\documentclass{elsart}
\usepackage{amssymb,amsmath,stmaryrd}
\usepackage[mathscr]{eucal}
\usepackage{epsfig}
\usepackage{pstricks,pst-node,pst-text,pst-3d}

\newcommand{\svee}{\operatornamewithlimits{\varovee}}
\newcommand{\swedge}{\operatornamewithlimits{\varowedge}}
\newcommand{\sign}{\mathrm{sign}\,}
\def \difw {\overset{\scriptscriptstyle{\mathrm{w}}}{\smash{-}}}

\def\1{{\mathchoice {\rm 1\mskip-4mu l} {\rm 1\mskip-4mu l}
{\rm 1\mskip-4.5mu l} {\rm 1\mskip-5mu l}}}
\def \0{{\mathbb{O}}}
\def \cho {\mathcal{C}}
\def \symcho {\check{\mathcal{C}}}
\def \sug {\mathcal{S}}
\def \symsug {\check{\mathcal{S}}}

\begin{document}
\begin{frontmatter}

\title{The M\"obius transform on  symmetric ordered structures and its
application to capacities on finite sets}

\author{Michel GRABISCH\thanksref{add}}
\thanks[add]{Mailing address: LIP6 ---
8, rue du Capitaine Scott, 75015 Paris, France}
\address{Universit\'e Paris I Panth\'eon-Sorbonne}
\ead{michel.grabisch@lip6.fr}

\begin{abstract}
Considering a linearly ordered set, we introduce its symmetric version, and
endow it with two operations extending supremum and infimum, so as to obtain an
algebraic structure close to a commutative ring. We show that imposing symmetry
necessarily entails non associativity, hence computing rules are defined in
order to deal with non associativity. We study in details computing rules,
which we endow with a partial order. This permits to find solutions to the
inversion formula underlying the M\"obius transform. Then we apply these
results to the case of capacities, a notion from decision theory which
corresponds, in the language of ordered sets, to order preserving mappings,
preserving also top and bottom. In this case, the solution of the inversion
formula is called the M\"obius transform of the capacity. Properties and
examples of M\"obius transform of sup-preserving and inf-preserving capacities
are given.
\end{abstract}
\end{frontmatter}

\section{Introduction}
We consider a linearly ordered set $(L^+,\leq)$, with bottom and top denoted by
$\0,\1$ respectively, and we define $L:=L^+\cup L^-$, where $L^-$ is a reversed
copy of $L^+$, i.e. for any $a,b\in L^+$, we have $a\leq b$ iff $-b\leq -a$,
where $-a,-b$ are the copies of $a,b$ in $L^-$. The set of signed integers, the
set of real numbers have this structure, with a central zero, and they possess
rich algebraic structures (groups, rings, etc.) when endowed with usual
arithmetical operations.

Our aim is to build similar structures, but using only the order relation
$\leq$ on $L$, so that the resulting structure should be as close as possible
to e.g. the ring of real numbers with $+,\times$. This should permit an easy
manipulation of functions or functionals (such as set functions, capacities,
integrals) taking values in $L$.

\bigskip

Our work is essentially motivated by decision making. Since this is fundamental
for our approach, we briefly introduce necessary notions. The aim of decision
making is to rank or assign overall scores to \emph{alternatives},
i.e. functions $f:S\longrightarrow L$, where $S$ is a set of features
(criteria, points of view, states of nature, etc.), and $f(s)\in L$ for any
$s\in S$ is the score of $f$ for feature $s$, expressed on some scale (usually
a real interval). Then overall scoring of $f$ can be viewed as a functional
$V:L^S\longrightarrow L$ satisfying certain properties. A very general way to
define $V$ is to take some integral. The Choquet integral \cite{cho53},
generalizing the Lebesgue integral, has proven to be a suitable and very
general functional for decision making \cite{sch86}, defined for non negative
functions. The Choquet integral is defined with respect to a \emph{capacity}
$v:2^S\longrightarrow [0,1]$, a monotone set function extending classical
measures used in the Lebesgue integral. For any capacity $v$, its
\emph{M\"obius transform} $m^v:2^S\longrightarrow \mathbb{R}$ is a key concept
in decision analysis (see e.g. \cite{chja89}). It is defined by
\begin{equation}
\label{eq:mobi}
m^v(A):=\sum_{B\subseteq A}(-1)^{|A\setminus B|}v(B), \quad\forall A\subseteq S.
\end{equation}
In a more general way, the M\"obius transform provides an inversion formula
useful in combinatorics \cite{rot64}.   

Let us denote by $\cho_v(f)$ the Choquet integral of $f$ with respect to
capacity $v$.  When $L$ happens to be a real interval containing negative
numbers, then a suitable extension of the Choquet integral has to be defined
for real-valued functions. It is called the \emph{symmetric Choquet integral},
and is defined as:
\begin{equation}
\label{eq:sipos}
\symcho_v(f) := \cho_v(f^+) - \cho_v(f^-)
\end{equation}
with $f^+=f\vee 0$, and $f^-=(-f)^+$. This is the basis for \emph{Cumulative
Prospect Theory} \cite{tvka92}, an important theory in economics for
representing human behaviour in decision making when faced with gains (positive
values, $L^+$) and losses (negative values, $L^-$).

If $L$ is only an \emph{ordinal scale}, i.e. a (usually finite) scale with only
a total order on it, then the Choquet integral is no more applicable, since
usual arithmetical operations are not defined on $L$. It is known that the
counterpart of it is the Sugeno integral \cite{sug74}, denoted $\sug_v(f)$,
which is defined solely with supremum ($\vee$) and infimum ($\wedge$), and like
the Choquet integral, with respect to a capacity $v$, which has to be valued on
$L$. However, in the ordinal case, there is no symmetric Sugeno integral, since
first of all there is no notion of ``negative numbers'' for ordinal
scales. Similarly, there is no M\"obius transform. Our aim is precisely to
define negative ordinal quantities so as to obtain a sufficiently rich
algebraic structure on $L$ to allow computation similar to (\ref{eq:mobi}),
(\ref{eq:sipos}), and thus to be able to develop an ordinal counterpart of
Cumulative Prospect Theory.

\bigskip

Generally speaking , we may think of several ways to tackle this problem. We
denote $\svee,\swedge$ the new operations on $L$.

An immediate solution would have been to use the Boolean ring associated to
$(L,\leq)$. But this solution works only if $(L,\leq)$ is a Boolean lattice,
and our application field requires that $L$ be only a linear lattice. 

A second solution is to define $\svee,\swedge$ as binary operators
$L^2\longrightarrow L$, and impose (possibly among other conditions) that
\begin{itemize}
\item [(\textbf{C1})] $\svee,\swedge$ coincide with $\vee,\wedge$ respectively 
on $L^+$
\item [(\textbf{C2})] $-a$ is the symmetric of $a$, i.e. $a\svee(-a) = \0$.
\item [(\textbf{C3})] $-(a\svee b)=(-a)\svee(-b)$, $-(a\swedge b)=(-a)\swedge b$. 
\end{itemize}
These conditions are motivated by our aim to develop an ordinal Cumulative
Prospect Theory:
\begin{enumerate}
\item (\textbf{C1}) permits us to perform an extension of all that already
exists in $L^+$, e.g. the Sugeno integral.
\item Thanks to (\textbf{C2}) and (\textbf{C3}),  computations
could be conducted as with real numbers, with $\svee,\swedge$ playing the role
of $+,\times$ respectively. In particular, it would permit to define a
counterpart of the M\"obius transform (\ref{eq:mobi}), and to define a
\emph{symmetric Sugeno integral}, in the spirit of (\ref{eq:sipos}):
\begin{equation}
\label{eq:symsug}
\symsug_v(f) := \sug_v(f^+) \svee(-\sug_v(f^-))
\end{equation}
Condition (\textbf{C2}) then implies that the integral of $f$ (overall scoring)
is $\0$ whenever $\sug_v(f^+)=  \sug_v(f^-)$, a desirable property since it
means that the overall scoring should be null when gains equal losses. 
\end{enumerate}
The problem with this solution is that due to (\textbf{C1}) and (\textbf{C2}),
inevitably $\svee$ would be non associative in general. Take $\0<a<b$ and
consider the expression $(-b)\svee b\svee a$. Depending on the place of
parentheses, the result differs since $((-b)\svee b)\svee a= \0\svee a = a$,
but $(-b)\svee(b\svee a)=(-b)\svee b=\0$. In other words, if we want to keep
associativity and (\textbf{C1}), then necessarily, (\textbf{C2}) does not hold:
Prop. \ref{prop:best} will show that in this case, $|a\svee(-a)|\geq
|a|$. Clearly, this result does not match intuition in our decision making
perspective, and hence we have no other way than to accept non associativity.
Remark however that as far as Eq. (\ref{eq:symsug}) is concerned, we need no
associativity property.

In order to escape the incompatibility between symmetry and associativity, a
third solution would be to define $\svee,\swedge$ as binary operators on
$(L^+\times L^-)^2\longrightarrow L^+\times L^-$, that is, each element $a$ in
$L$ is viewed as a pair $(a^+,a^-)\in L^+\times L^-$, where $a^-=\0$ if $a\geq
\0$, and $a^+=\0$ otherwise. Then for any $(a^+,a^-), (b^+,b^-)$ in $L^+\times
L^-$, one could define in an obvious way:
\begin{align*}
-(a^+,a^-):= & (-a^-,-a^+)\\
(a^+,a^-)\svee(b^+,b^-) : = & (a^+\vee b^+, a^-\wedge b^-)\\
(a^+,a^-)\swedge(b^+,b^-) : = & (a^+\wedge b^+, a^-\vee b^-).
\end{align*}
Thus $\svee,\swedge$ are associative since $\vee,\wedge$ are on $L^+,L^-$, they
coincide with $\vee,\wedge$ on $L^+$ (condition (\textbf{C1})), and condition (\textbf{C3}) is
fulfilled since $-((a^+,a^-)\svee(b^+,b^-)) =
(-(a^+,a^-))\svee(-(b^+,b^-))$. However, elements have no opposite,
since\linebreak 
$(a^+,a^-)\svee(-a^-,-a^+)\neq(\0,\0)$. Also, there is no total order on
$L^+\times L^-$.

Considering our motivation, only the second solution is acceptable, since
symmetry is mandatory in our framework, and the third solution
would lead to a partial order on alternatives, a situation which is not
desirable in decision making.

In this paper, our aim is to completely develop the second solution, and to
apply it in particular to the definition of an ordinal M\"obius transform, the
definition of the symmetric Sugeno integral being already solved as indicated
above. 
First attempts at defining the M\"obius transform of capacities in an ordinal
context have been done by Marichal \cite{mamato96,mar00a}, Mesiar \cite{mes97},
and the author \cite{gra97a}. However, these preliminary works have been done
without explicit connection to combinatorics, and were restricted to
capacities.

\bigskip

The paper is organized as follows. Next section introduces necessary concepts
for the sequel, while Section 3 gives the construction of the symmetric ordered
structure. Since $\svee$ is necessarily non associative, Section 4 introduces
rules of computation, which give meaning to expression such as $\svee_{i\in
I}a_i$. Section 5 is devoted to the study of the M\"obius inversion formula,
when defined on symmetric ordered structures. Lastly, Section 6 focusses on
capacities, while Section 7 concludes the paper by indicating possible
applications of the results. 

\section{Preliminaries}
We give necessary definitions and introduce basic concepts for our
construction. 

Let $N$ be a finite set, and $(L^+,\leq)$ a totally ordered set, with $\0,\1$
being top and bottom elements. A ($L^+$-valued) \emph{capacity} is an order
preserving (or \emph{isotone}) mapping $v:(2^N, \subseteq)\longrightarrow
(L^+,\leq)$, with $v(\emptyset) = \0$, $v(N) = \1$. 

\bigskip

We say that a complete lattice $(\mathcal{L},\leq)$ is a \emph{conjugation}
lattice if it is endowed with a bijective and order-reversing mapping from
$\mathcal{L}$ to $\mathcal{L}$, called a \emph{conjugation}, which maps $a$ to
$\overline{a}$, so that $\overline{\overline{a}}=a$, and $a\leq b$ iff
$\overline{a}\geq\overline{b}$. Then $\overline{a\vee
b}=\overline{a}\wedge\overline{b}$, and $\overline{a\wedge
b}=\overline{a}\vee\overline{b}$. In the Boolean lattice $2^N$, set complement
is a conjugation.

If $(L^+,\leq)$ has a conjugation, then the \emph{conjugate capacity}
$\overline{v}$ is defined by $\overline{v}(A) := \overline{v(\overline{A})},
\quad A\subseteq N$.

\bigskip

Let us consider a poset $(P^+,\leq)$, with bottom and top
elements denoted by $\0$ and $\1$. We introduce $P^-:=\{-a|a\in P^+\}$, with
the reversed order, i.e. $-a\leq -b$ iff $a\geq b$ in $P^+$. The bottom and top
of $P^-$ are respectively $-\1$ and $-\0$. 

The disjoint union of $P^+$ and $P^-$, with identification of $-\0$ with $\0$,
is called a \emph{reflection poset} or \emph{symmetric poset}, and is denoted
$(P,\leq)$ \cite{degr01a}. It is a poset with bottom $-\1$ and top $\1$.

We introduce some mappings on $(P,\leq)$. The \emph{reflection} maps $a\in P$
to $-a$, and $-(-a):=a$ for any $a\in P$. If $P$ is a lattice we have:
\[
(-a)\vee(-b) = -(a\wedge b), \quad (-a)\wedge(-b) = -(a\vee b).
\]
The \emph{absolute value} of $a\in P$ is denoted $|a|$, and $|a|:=a$ if $a\in
P^+$, and $|a|=-a$ otherwise. The \emph{sign function} is defined by:
\[
\sign : P \rightarrow P     
\,,\quad \sign x=\left\{ \begin{array}{ll}
-{\1} & \mbox{ for } x < {\0}\\
{\0} & \mbox{ for } x = {\0}\\
{\1} & \mbox{ for } x > {\0}
\end{array}
\right. \,.
\]

Lastly, we introduce the notion of derivative.
\begin{defn}
\label{def:deri}
Let $(X,\leq)$ be a poset and $(\mathcal{L},\leq)$ be a complete lattice.
For any isotone function $g$ from $X$ to $\mathcal{L}$, let
$\tilde{g}$ be defined by:
\begin{equation}
\tilde{g}(x) := \bigvee_{y<x}g(y).
\end{equation}
Then the \emph{derivative} $g'$ of $g$ is defined by:
\[
g'(x):= \left\{ \begin{array}{ll}
                        \0, & \text{ if } g(x)=\tilde{g}(x),\\
                        g(x), & \text{ otherwise}.
                                                \end{array}     \right.
\] 
\end{defn}

In a partially ordered set (poset for short) $(X,\leq)$, we say that $x$
\emph{covers} $y$, denoted by $x\succ y$, if $x> y$, and there is no $u\in
X$ such that $u\neq x,y$ and $x>u>y$.

\section{Symmetric ordered structures}
Let $(L^+,\leq)$ be a totally ordered set (linear lattice) with top and bottom
$\1,\0$, and consider the corresponding symmetric linear lattice $(L,\leq)$. As
stated in the introduction, our aim is to endow $L$ with two operations
denoted $\svee,\swedge$ extending usual $\vee,\wedge$ on $L^+$, so that the
resulting structure is close to a ring. More precisely, we require the
following:
\begin{itemize}
\item [(\textbf{C1})] $\svee,\swedge$ coincide with $\vee,\wedge$ respectively on $L^+$
\item [(\textbf{C2})] $-a$ is the symmetric of $a$, i.e. $a\svee(-a) = \0$,
$\forall a\in L$.
\item [(\textbf{C3})] $-(a\svee b)=(-a)\svee(-b)$, $-(a\swedge b)=(-a)\swedge b$,
$\forall a,b\in L$. 
\end{itemize}
Let us build $\svee$ first. Due to (\textbf{C1}), $\svee$ is defined on $L^+$ and
coincides with $\vee$, which implies that $a\svee\0=a$ for any $a>\0$. Using
(\textbf{C3}), we deduce $-(a\svee\0)=(-a)\svee\0=-a$, showing that $\0$ is the neutral
element. Again using (\textbf{C3}) with $a,b\geq\0$, we get
\[
-(a\svee b)=-(a\vee b)=(-a)\wedge(-b)=(-a)\svee(-b)
\]
showing that $\svee$ coincides with $\wedge$ on $L^-$. It remains to define
$\svee$ for arguments of opposite sign. We propose the following (this will be
justified in Prop. \ref{prop:best}), assuming $a>\0$ and $b<\0$
\begin{equation}
\label{eq:symdif}
a\svee b :=\begin{cases}
                a, & \text{ if } a>-b\\
                \0, & \text{ if } a=-b\\
                b, & \text{ otherwise.}
                \end{cases}
\end{equation}
Note that the second case is just (\textbf{C2}). Using (\textbf{C3}), we can
derive the formula for the opposite case $a<\0$ and $b>\0$. In summary, $\svee$
is given by Fig. \ref{tab:vvee}.
\begin{figure}[htb]
\begin{center}
\psset{unit=1cm}
\pspicture(-0.5,-0.5)(4.5,4.5)
\psline{->}(-0.5,2)(4.5,2)
\psline{->}(2,-0.5)(2,4.5)
\pspolygon(0,0)(0,4)(4,4)(4,0)
\psline[linewidth=2pt](0,4)(4,0)
\pscurve{->}(0.5,4.2)(1.25,3.75)(1,3)
\pscurve{->}(4.2,1.5)(3.8,1.4)(3.5,0.5)
\uput[180](0.5,4.2){$\0$}
\uput[0](4.2,1.5){$\0$}
\uput[0](4.5,2){$a$}
\uput[0](2,4.5){$b$}
\uput[45](4,2){\small $\1$}
\uput[135](0,2){\small $-\1$}
\uput[45](2,2){\small $\0$}
\uput[135](2,4){\small $\1$}
\uput[225](2,0){\small $-\1$}
\rput(3,3){$a\vee b$}
\rput(1,1){$a\wedge b$}
\rput(1.5,3.5){$b$}
\rput(2.5,0.5){$b$}
\rput(3.5,1.5){$a$}
\rput(0.5,2.5){$a$}
\endpspicture
\end{center}
\caption{Definition of the symmetric maximum $a\varovee b$}
\label{tab:vvee}
\end{figure}

A compact formulation of $\svee$ is:
\begin{equation}
\label{eq:symmax}
a \svee b := \left\{ \begin{array}{ll}
-(|a| \vee |b|) & \mbox{ if } b \neq -a \mbox{ and } |a| \vee |b|=-a \mbox{ or } =-b
\\
{\0} & \mbox{ if } b=-a \\
|a| \vee |b| & \mbox{ else.}
\end{array}
\right.
\end{equation}
Except for the case $b=-a$, $a \svee b$ equals the larger one in absolute value
of the two elements $a$ and $b$.
\begin{rem}
The following interpretation can be given for $\svee$: on scale $L$, distinct
levels are far away from one another, so that invoking negligibility aspects,
only the maximum level remains when combining two positive values. When a
positive (gain) and a negative value (loss) are combined, if the gain dominates
the loss, the latter counts for nothing. 
\end{rem}
\begin{rem}
Equation (\ref{eq:symdif}) is a symmetrized version of a difference operator
introduced by Weber \cite{web84}:
\begin{equation}
a\difw b := \inf\{c|b\vee c\geq a\} = \left\{\begin{array}{ll}
                                                a, & \text{ if } a>b\\
                                                \0, & \text{ otherwise}
                                                \end{array} \right.
\end{equation}
for any $a,b\in L^+$. Note that $a\difw b$ is the dual of the pseudo-complement
of $b$ relative to $a$, defined by $b*a:=\sup\{c|b\wedge c\leq a\}$ (see
e.g. \cite{gratz98}). It is also the dual of the residual of $a$ by $b$
(see e.g. \cite{bir67}).
\end{rem}

It remains to define the \emph{symmetric minimum} operator. Since we impose the
symmetry condition (\textbf{C3}), we are naturally lead to
Fig \ref{tab:vwedge}. 
\begin{figure}[htb]
\begin{center}
\psset{unit=1cm}
\pspicture(-0.5,-0.5)(4.5,4.5)
\psline{->}(-0.5,2)(4.5,2)
\psline{->}(2,-0.5)(2,4.5)
\pspolygon(0,0)(0,4)(4,4)(4,0)
\uput[0](4.5,2){$a$}
\uput[0](2,4.5){$b$}
\uput[45](4,2){\small $\1$}
\uput[135](0,2){\small $-\1$}
\uput[45](2,2){\small $\0$}
\uput[135](2,4){\small $\1$}
\uput[225](2,0){\small $-\1$}
\rput(3,3){$a\wedge b$}
\rput(1,1){$|a|\wedge |b|$}
\rput(1,3){$-(|a|\wedge b)$}
\rput(3,1){$-(a\wedge |b|)$}
\endpspicture
\end{center}
\caption{Definition of the symmetric minimum $a\varowedge b$}
\label{tab:vwedge}
\end{figure}

A more compact expression is:
\begin{equation}
a \swedge b := \left\{ \begin{array}{ll}
-(|a| \wedge |b|) & \mbox{ if } \sign a \neq \sign b \\
|a| \wedge |b| & \mbox{ else.}
\end{array}
\right. 
\end{equation}
The absolute value of $a \swedge b$ equals $|a| \wedge |b|$ and $a \swedge b <
{\0}$ iff the two elements $a$ and $b$ have opposite signs.

Another equivalent formulation of these two operations, applicable if $L$ is a
symmetric real interval, is due to Marichal \cite{mar01}, and clearly shows the
relationship with the ring of real numbers.
\begin{align}
a\svee b & = \sign(a+b)(|a|\vee |b|)\\
a\swedge b & = \sign(a\cdot b)(|a|\wedge|b|).
\end{align}

The following proposition summarizes the properties of the structure obtained. 
\begin{prop}
\label{prop:ring}
The structure $(L,\varovee,\varowedge)$ has the
following properties.
\begin{itemize}
\item [(i)] $\varovee$ is commutative.
\item [(ii)] $\0$ is the unique neutral element of $\varovee$, and the
unique absorbing element of $\varowedge$.
\item [(iii)] $a\varovee -a=\0$, for all $a\in L$.
\item [(iv)] $-(a\varovee b) = (-a)\varovee (-b)$.
\item [(v)] $\varovee$ is associative for any expression involving
$a_1,\ldots,a_n$, $a_i\in L$, such  that $\bigvee_{i=1}^n a_i\neq
-\bigwedge_{i=1}^n a_i$.
\item [(vi)] $\varowedge$ is commutative.
\item [(vii)] $\1$ is the unique neutral element of $\varowedge$, and the
unique absorbing
element of $\varovee$.
\item [(viii)] $\varowedge$ is associative on $L$.
\item [(ix)] $\varowedge$ is  distributive w.r.t $\varovee$ in
$L^+$ and $L^-$ separately.
\item [(x)] $\svee$ is isotone, i.e. $a\leq a', b\leq b'$ implies $a\svee b\leq
a'\svee b'$.
\end{itemize}
\end{prop}
\begin{pf}
All results are almost clear from the construction. We just detail (v) and
(ix). 

(v) Let us study if the equality $(a\svee b)\svee c = a\svee(b\svee c)$ holds
supposing there is no pair of symmetric elements, as $(a,-a)$. This implies
$|a\svee b|=|a|\vee|b|$ (see (\ref{eq:symmax})). Hence
\[
|(a\svee b)\svee c| = |a\svee b|\vee|c| = |a|\vee|b|\vee|c|=|a|\vee|b\svee c| =
 |a\svee(b\svee c)|.
\]
Thus, $(a\svee b)\svee c$ and $a\svee(b\svee c)$ have the same absolute
value. It remains to prove that they have the same sign. The sign of $a\svee b$
is the sign of the largest term in absolute value. Hence, the sign of $(a\svee
b)\svee c$ is the sign of the largest in absolute value among $a\svee b$ and
$c$, so it is the sign of the largest in absolute value among $a,b,c$. Doing
the same with $a\svee(b\svee c)$, we conclude that the two expressions have the
same sign.

Suppose now $a=-b$. Then $(a\svee b)\svee c=c$. Clearly, $a\svee(b\svee c)=c$
if and only if $|c|>|a|$. This coincides with the condition given in (v). 

(ix) Distributivity is clearly satisfied on $L^+$. For any $a,b,c\in L^-$:
\begin{align*}
(a\svee b)\swedge c & = (a\wedge b)\swedge |c| = |a\wedge b|\wedge |c|\\
(a\swedge c)\svee(b\swedge c) & = (|a|\wedge |c|)\vee(|b|\wedge |c|)\\
        & = (|a|\vee|b|)\wedge |c|\\
        & = |a\wedge b|\wedge |c|.
\end{align*}
\qed
\end{pf}

The distributivity does not hold in general: take $a,b\geq \0$,
$a<b$, $c<\0$, $b<-c$. Then 
\begin{align*}
a\varowedge (b\varovee c) & = a\varowedge c = -a\\
(a\varowedge b)\varovee(a\varowedge c) & = a\varovee(-a) = \0.
\end{align*} 

Using the definition of the symmetric maximum, we see that the derivative of a
function $g$ (see Definition \ref{def:deri}) can be reformulated as:
\begin{equation}
\label{eq:deriv}
g'(x) = g(x)\svee (-\tilde{g}(x)).
\end{equation}

The next proposition gives justifications to our choice in (\ref{eq:symdif}),
and of the overall construction. 
\begin{prop}
\label{prop:best}
We consider conditions (\textbf{C1}), (\textbf{C2}) and (\textbf{C3}), and
denote by (\textbf{C3+}) condition (\textbf{C3}) when $a,b$ are restricted to
$L^+$. Then:

(0) Conditions (\textbf{C1}) and (\textbf{C2}) implies that associativity cannot hold.

(1) Under conditions (\textbf{C1}), (\textbf{C2}) and (\textbf{C3}), no operation is associative on a
larger domain than $\svee$ as given by (\ref{eq:symmax}).

(2) Under (\textbf{C1}) and (\textbf{C3+}), $\0$ is neutral. If we require in addition
    associativity, then $|a\svee(-a)|\geq |a|$. Further, if we require
    isotonicity of $\svee$, then  $|a\svee(-a)|=|a|$.
\end{prop}
\begin{pf}
(0) Let us take $\0<a<b$. Then $((-b)\svee b)\svee a=\0\svee
    a=a\neq(-b)\svee(b\svee a)= (-b)\svee b=\0$ (see introduction).  

(1) The only degree of freedom is the definition of
$a\svee b$ when $a,b$ have different signs. We know that
the only non associative case happens in expressions
like $-x\svee(x\svee y)$, $x,y\geq\0$. Since $\svee\equiv\vee$ on $[\0,\1]^2$,
we get:
\[
-x\svee(x\svee y)= -x\svee(x\vee y)= \left\{\begin{array}{ll}
                                -x\svee x=\0, & \text{ if } x\geq y\\
                                -x\svee y, & \text{ if } x\leq y.
                                \end{array}\right.
\]
Observing that $(-x\svee x)\svee y = y$, clearly the first case can never
lead to associativity. Let us examine the second case. It leads to
associativity iff $-x\svee y=y$. Discarding the case $x=y$, we see that we
have in fact the definition of the symmetric maximum. Hence only it can lead to
associativity in this case, and any other operation would not. 

(2) Let us assume (\textbf{C1}) and (\textbf{C3+}). If $a>\0$, then
    $a\svee\0=a$, and $-(a\svee\0)=(-a)\svee\0=-a$. Now, if associativity
    holds, then taking $a>\0$, we have $((-a)\svee a)\svee a = (-a)\svee(a\svee
    a)$, which gives $((-a)\svee a)\svee a = (-a)\svee a$. We know from (0)
    that (\textbf{C1}) and associativity imply that $(-a)\svee a\neq\0$. If
    $(-a)\svee a>\0$, then to satisfy the above equality we must have
    $(-a)\svee a\geq a$. If it is a negative, a similar argument using
    (\textbf{C3+}) shows that $(-a)\svee a\leq -a$. Lastly, if $\svee$ is
    isotone, we have $a\svee(-a)\leq a\svee\0=a$, and similarly for the
    negative case.
\qed
\end{pf}

\section{Non associativity and computing rules}
\label{sec:nonass}
Due to the lack of associativity of $\svee$, expressions like $\svee_{i=1}^n
a_i$ have no meaning, unless one defines a particular and systematic way of
arranging terms so that associativity problems disappear. 

Let us consider a sequence $\{a_i\}_{i\in I}$ of terms $a_i\in L$, with
$I\subseteq \mathbb{N}$. We say that the sequence \emph{fulfills associativity}
if either $|I|\leq 2$ or $\vee_{i\in I}a_i\neq\wedge_{i\in I}a_i$. Hence, from
Prop. \ref{prop:ring} (v), $\svee_{i\in I}a_i$ is well-defined if and only if
the sequence $\{a_i\}_{i\in I}$ is associative. If a sequence does not fulfill
associativity, it necessarily has at least 3 terms and contains a pair of
maximal opposite terms $(a,-a)$, with $a:=\vee_{i\in I}a_i$. Discarding all
occurrences of $a,-a$ in the sequence, we may still find (new) maximal opposite
terms $b,-b$, which can be discarded, etc., until no more such terms remain,
which means that the new sequence fulfills associativity. We call the
\emph{sequence of maximal opposite terms} the sequence of all deleted terms,
whose index set is denoted $I_=$. Taking for example with $L=\mathbb{Z}$ the
sequence $3, 3, 3, 2, 1, 0, -2, -3, -3$, the sequence of maximal opposite terms
is $3,3,3,2,-2,-3,-3$.

Another way to fulfill associativity is obtained by discarding in the sequence
the pair $(a,-a)$, with $a:=\vee_{i\in I}a_i$, and if the new sequence
$\{a_i\}_{i\in I}\setminus \{a,-a\}$ does not fulfill associativity, then
discard the pair of maximal opposite terms in this new sequence, etc., until
associativity is fulfilled. We call the \emph{restricted sequence of maximal
opposite terms} the sequence of all deleted terms, and we denote its index set
by $I_0$. In the previous example, the restricted sequence of maximal opposite
terms is $3,-3,3,-3$. Note that we always have $I_0 \subseteq I_=$, and that
$I_0$ is minimal in the sense that no proper subset of it can ensure
associativity.

We denote the set of all (at most countable) sequences, including the empty
one, by $\mathfrak{S}:={\displaystyle \bigcup_{i=1}^\infty
L^i\cup\{\emptyset\}}$.  From now on, we make the convention $\svee_\emptyset
a_i=\0$.

A \emph{computation rule} is a systematic way to delete terms in a sequence
$\{a_i\}_{i\in I}$, so that it fulfills associativity, provided the way they
are deleted can be obtained as the result of a suitable arrangement of
parentheses in $\svee_{i\in I}a_i$. For example, deleting 3 in the sequence
$3,1,-3$ makes the sequence associative, but does not correspond to some
arrangement of parentheses, and so is not a computation rule. Formally, we
denote a computation rule by the infix notation:
\[
\langle\cdot\rangle:\begin{array}{lcl}
        \mathfrak{S}&\longrightarrow & \mathfrak{S}\\
        \{a_i\}_{i\in I} & \mapsto & \langle\{a_i\}_{i\in
        I}\rangle:=\{a_i\}_{i\in I\setminus J}
                \end{array}
\] 
where $J\subseteq I$ is the index set of deleted terms. To avoid heavy
notations, we denote $\svee_{i\in I\setminus J}\langle\{a_i\}_{i\in
I}\rangle$ by $\langle\svee_{i\in I}a_i\rangle$.  The set of all computation
rules on $L$ is denoted by $\mathfrak{R}$. 

Let us give some basic examples of computation rule.
\begin{enumerate}
\item The \emph{weak rule} $\langle\cdot\rangle_=$, where the index set of
deleted terms is $J=I_=$. It obviously corresponds to a particular arrangement
of parentheses, as shown in the following example:
\begin{multline}
\langle 3\svee 3\svee 3\svee 2\svee 1\svee 0\svee -2\svee -3\svee -3\rangle_=
=\\ ((3\svee 3\svee 3)\svee(-3\svee -3))\svee(2\svee -2)\svee(1\svee 0)=1.
\end{multline}
\item The \emph{strong rule} $\langle\cdot\rangle_0$, whose index set of
deleted terms is $I_0$. It obviously corresponds to a particular arrangement
of parentheses. Our example gives
\begin{multline}
\langle 3\svee 3\svee 3\svee 2\svee 1\svee 0\svee -2\svee -3\svee -3\rangle_0
=\\ (3\svee-3)\svee(3\svee -3)\svee(3\svee 2\svee 1\svee 0\svee -2)=3.
\end{multline}
\item The \emph{splitting rule} $\langle\cdot\rangle_-^+$, whose index set of
deleted terms is $J=\emptyset$ if the sequence fulffills associativity, and
$J=I$ if not. Then in the latter case, $\langle\svee_{i\in I}
a_i\rangle_-^+=\0$, due to our convention $\svee_\emptyset a_i=0$. The
corresponding arrangement of parentheses is 
\[
\langle\svee_{i\in I} a_i\rangle_-^+ := \Big(\svee_{a_i\geq
\0}a_i\Big)\svee\Big(\svee_{a_i< \0}a_i\Big).
\]
hence the name of the rule (splitting positive and negative terms). 
\item The \emph{optimistic} and \emph{pessimistic} rules
$\langle\cdot\rangle_\mathrm{opt}$, $\langle\cdot\rangle_\mathrm{pes}$. Let us
consider a sequence of at least 3 terms in $\mathfrak{S}$ having maximal
opposite terms $a,-a$, with degrees of multiplicity $k_+,k_-$ respectively. If
$k_+=1$ and $k_-\leq 2$, or $k_-=1$ and $k\leq 2$, then $J=I$ for both rules
(hence they give $\0$ as result). Otherwise, the optimistic rule deletes
$k_+-1$ occurrences of $a$ and all $k_-$ occurrences of $-a$ (hence it returns
$a$), while the pessimistic rule deletes all $k_+$ occurrences of $a$ and
$k_--1$ occurrences of $-a$ (hence it returns $-a$). One can verify that these
rules can be expressed as a particular arrangement of parentheses. For example
\begin{multline}
\langle 3\svee 3\svee 3\svee 2\svee 1\svee 0\svee -2\svee -3\svee -3\rangle_\mathrm{pes} =\\
((3\svee 3\svee 3)\svee -3)\svee (-3\svee 2\svee -2\svee 1\svee 0)=-3.
\end{multline}
\end{enumerate}
\begin{rem}
The first three rules have a clear meaning in decision making. Assume that
$\{a_i\}_{i\in I}$ is a sequence of scores assigned to some alternative. The
quantity $\svee_{i\in I}a_i$ is the overall score of the alternative. If the
splitting rule is used, the overall score is $\0$ whenever best and worse
scores are opposite. This way of computing the overall score is not vey
discriminating since many alternatives will get $\0$ as overall score, even if
the scores assigned to them are very different. The two other rules are more
discriminating since they discard maximal opposite scores: if best and worst
scores are opposite, then look at second best and second worst scores, etc.

The purpose of the optimistic and pessimistic rules are merely for illustration
of properties. They obviously have no ``rational'' behaviour in a decision
making framework.
\end{rem}
\begin{rem}
The  strong rule coincides with the limit of some
family of uni-norms proposed by Mesiar and Komornikov\'a \cite{meko98}
(uni-norms are binary operations on $[0,1]^2$ which are associative,
commutative, non decreasing and with a neutral element $e\in ]0,1[$. See
\cite{klmepa00} for details). 
\end{rem}

Let us endow $\mathfrak{R}$ with the following order: for
$\langle\cdot\rangle_1,\langle\cdot\rangle_2\in\mathfrak{R}$,
$\langle\cdot\rangle_1\sqsubseteq\langle\cdot\rangle_2$ iff for all sequences
$\{a_i\}_{i\in I}\in\mathfrak{S}$, $J_1\supseteq J_2$, where $J_1,J_2$ are the
index sets of deleted terms for rules 1 and 2. $\sqsubseteq$ being reflexive,
antisymmetric (since computation rules are precisely defined by the set of
deleted terms) and transitive, $(\mathfrak{R},\sqsubseteq)$ is a partially
ordered set. As usual, the interval
$[\langle\cdot\rangle_1,\langle\cdot\rangle_2]$ denotes the set of all
computation rules $\langle\cdot\rangle$ such that
$\langle\cdot\rangle_1\sqsubseteq\langle\cdot\rangle\sqsubseteq\langle\cdot\rangle_2$.

Let us introduce another order relation on $\mathfrak{R}$.  The sequence
$\{a_i\}_{i\in I}$ in $\mathfrak{S}$ is said to be a \emph{cancelling sequence
for the rule $\langle\cdot\rangle$} if $\langle\svee_{i\in I} a_i\rangle=\0$.
We denote by $\mathcal{O}_{\langle\cdot\rangle}$ the set of cancelling
sequences of $\langle\cdot\rangle$. We say that computation rule
$\langle\cdot\rangle_1$ is more \emph{discriminating} than rule
$\langle\cdot\rangle_2$, denoted as
$\langle\cdot\rangle_1\succcurlyeq\langle\cdot\rangle_2$, if
$\mathcal{O}_{\langle\cdot\rangle_1}\subseteq
\mathcal{O}_{\langle\cdot\rangle_2}$. Note that $\succcurlyeq$ is only a
preorder, since being reflexive and transitive, but not antisymmetric.  For a
justification of the name ``discriminating'', see Remark~3
\begin{prop}
\label{prop:cr}
For any rules $\langle\cdot\rangle_1,\langle\cdot\rangle_2\in\mathfrak{R}$, the
following holds:
\begin{itemize}
\item [(i)]$\langle\cdot\rangle_1\sqsubseteq\langle\cdot\rangle_2$ implies that
for all sequences $\{a_i\}_{i\in I}\in\mathfrak{S}$, $|\langle\svee_{i\in
I}a_i\rangle_1|\leq |\langle\svee_{i\in I}a_i\rangle_2|$.
\item [(ii)]$\langle\cdot\rangle_1\sqsubseteq\langle\cdot\rangle_2$
implies
$\mathcal{O}_{\langle\cdot\rangle_1}\supseteq\mathcal{O}_{\langle\cdot\rangle_2}$
\item [(iii)]$\langle\cdot\rangle^+_-$ is the unique minimal element of
$(\mathfrak{R},\sqsubseteq)$, while $\langle\cdot\rangle_0$ is a maximal element. 
\end{itemize}
\end{prop}
\begin{pf}
(i) Let $\{a_i\}_{i\in I}$ not fulfilling associativity. Rule $k$, $k=1,2$,
makes the sequence associative by removing terms $a_i$, $i\in J_k$. Then
$\langle\svee_{i\in I\setminus J_k}a_i\rangle_k$ equals either $\vee_{i\in
I\setminus J_k} a_i$ or $\wedge_{i\in I\setminus J_k}a_i$. By hypothesis,
$J_1\supseteq J_2$, hence the result.

(ii) Let $A$ be a cancelling sequence for rule 2, which means that
$\langle\svee_{a\in A}a\rangle_2=\0$. Then applying (i), clearly $A$ is a
cancelling sequence for rule 1. 

(iii) Obvious for $\langle\cdot\rangle^+_-$.  $\langle\cdot\rangle_0$ is a
maximal element since the sequence of deleted terms is $I_0$, which is a
minimal sequence as remarked above (no proper subset can ensure associativity).
\qed
\end{pf}

The following corollary is immediate. 
\begin{cor}
\begin{itemize}
\item [(i)] for any sequence in $\mathfrak{S}$, $|\langle\svee_{i\in
I}a_i\rangle_0|\geq|\langle\svee_{i\in I}a_i\rangle_=|\geq|\langle\svee_{i\in
I}a_i\rangle_-^+|$, and $|\langle\svee_{i\in I}a_i\rangle_-^+|$ is the lowest
bound of $|\langle\svee_{i\in I}a_i\rangle|$ for all rules
$\langle\cdot\rangle$ in $\mathfrak{R}$.
\item [(ii)] $\mathcal{O}_{\langle\cdot\rangle_0}\subseteq
\mathcal{O}_{\langle\cdot\rangle_=}\subseteq
\mathcal{O}_{\langle\cdot\rangle_-^+}$.
\end{itemize}
\end{cor}
$(\mathfrak{R},\sqsubseteq)$ fails to be a lattice or even a semi-lattice, as
shown by the following example. Consider the optimistic and pessimistic rules
and the following sequence: $3, 3, 3, 2,1, -2, -3, -3,-3$. The terms deleted by
these rules are $J_\mathrm{opt}=3, 3, -3, -3, -3$, and $J_\mathrm{pes}=3, 3, 3,
-3, -3$. An upper bound of the two rules deletes at most the terms in
$J_\mathrm{opt}\cap J_\mathrm{pes}=3,3,-3,-3$. In any case, the resulting
sequence is not associative, hence it does not define a computation
rule. Similarly, a lower bound deletes at least $J_\mathrm{opt}\cup
J_\mathrm{pes}=3,3,3,-3,-3,-3$. It is easy to see that $3,3,3,2,-3,-3,-3$ and
$3, 3,3,-2,-3,-3,-3$ are two maximal lower bounds each defining a computation
rule, hence there is no greatest lower bound.

We give hereafter some other properties. 
\begin{prop}
\label{prop:cr2}
For any sequences $\{a_i\}_{i\in I}$, $\{a'_i\}_{i\in I}$, and $\{b_i\}_{i\in
J}$ in $\mathfrak{S}$
\begin{itemize}
\item [(i)]
\[
\bigwedge_{i\in I} a_i\leq \langle\svee_{i\in I} a_i\rangle\leq \bigvee_{i\in
I} a_i .
\]
\item [(ii)] The rules $\langle\cdot\rangle_-^+$ and $\langle\cdot\rangle_0$
are isotone, i.e. they satisfy
\[
a_i\leq a'_i, \quad \forall i\in I\text{ implies } \langle \svee_{i\in I}
a_i\rangle \leq \langle\svee_{i\in I} a'_i\rangle.
\] 
\item [(iii)] \mbox{}
\begin{gather*}
|\langle(\svee_{i\in I} a_i) \svee(\svee_{j\in J} b_j)\rangle_-^+|\geq
 |\langle\svee_{i\in I} a_i\rangle_-^+|\\
\text{or } \langle(\svee_{i\in I} a_i) \svee(\svee_{j\in J} b_j)\rangle_-^+=\0.
\end{gather*}
\end{itemize}
\end{prop}
\begin{pf}
(i) Clear from definition.

(ii) It suffices to show the result for one argument, say $a_j$. Let us consider the
rule $\langle\cdot\rangle_-^+$. If $a_j\geq \0$, then by Prop. \ref{prop:ring}
(x), $\svee_{a_i\geq \0}a_i$ will not decrease when $a_j$ is replaced by
$a'_j$, so that $\svee_{i\in I}a_i$ will not decrease too (similarly if
$a_j<\0$).

We turn to the rule $\langle\cdot\rangle_0$. We consider the sequence
$\{a_i\}_{i\in I}$, and the index set of deleted terms $J$. If $j \in
I\setminus J$, then the expression $\svee_{i\in I\setminus J}a_i$ is
isotone provided associativity still holds when $a_j$ is replaced by $a'_j$
(see Prop. \ref{prop:ring} (x)). Since $a'_j\geq a_j$, the only case where
associativity is lost is when $\svee_{i\in I\setminus J}a_i=a_k$ with
$a_k<\0$, and $a'_j=-a_k$. In this case $a_k,a'_j$ are deleted,
and the result is the 2nd largest in absolute value, which is greater or equal
to $a_k$, hence the rule is still isotone.

Let us consider the case when $j\in J$, and suppose that $\svee_{i\in
I\setminus J}a_i=a_k $. If $a_j>\0$, then for $a'_j>a_j$, the pair
$(a'_j,-a_j)$ is no more deleted, and the result of computation will be
$a'_j$. Since $a'_j>a_j\geq a_k$, the rule is isotone. Now, if $a_j<\0$, for
$a'_j>a_j$, the pair $(a'_j,-a_j)$ is no more deleted, and the result
becomes $-a_j$. Since $-a_j\geq a_k$, isotonicity holds in this case too.

(iii) Let us consider a sequence $\{a_i\}_{i\in I}$, not fulfilling
associativity. We need only to prove the result for a sequence $\{b_j\}_{j\in
J}$ reduced to a singleton $b_1$, the general case follows by induction. We
denote $a:=\langle\svee_{i\in I} a_i\rangle^+_-$, and $b:=\langle(\svee_{i\in
I} a_i) \svee b_1\rangle_-^+$. We have $a=a^+\svee a^-$, with
$a^+:=\svee_{a_i\geq \0}a_i$, $a^-:=\svee_{a_i< \0}a_i$. Assume that
$a=a^+$. If $b_1\geq\0$, we have $b=a^+\vee b_1\geq a$. If $b_1<\0$, $b=a$
unless $b_1\leq -a^+$. If $b_1=-a^+$, then $b=\0$, and if $b_1<-a^+$, then
$b=b_1$, so that $|b|>|a|$. Assume now that $a=\0$, then trivially the result
holds. The case where $a=a^-$ works similarly as the case $a=a^+$.
\qed
\end{pf}

Computation rule $\langle\cdot\rangle_=$ is not isotone, as shown by the
following example: take the sequence $-3,3,1$ in $\mathbb{Z}$. Applying the
weak rule leads to 1. Now, if 1 is raised to 3, the result becomes $0$.   

\section{The ordinal M\"obius transform}
\label{sec:motr}
Throughout this section, let $(X,\leq)$ denote a locally finite poset (i.e. any
segment $[u,v]:=\{x\in X|u\leq x\leq v\}$ is finite) with unique minimal
element 0. We begin by briefly recalling the classical construction of the
M\"obius transform (see e.g. \cite{ber71,rot64}), and its connection with
capacities.

\subsection{Basic facts on the M\"obius transform}
Let us consider $f,g$ two real-valued functions on $X$ such that
\begin{equation}
\label{eq:mob1}
g(x) = \sum_{y\leq x} f(y).
\end{equation}
A fundamental question in combinatorics is to solve this equation, i.e. to
recover $f$ from $g$. The solution is given through the \emph{M\"obius
function} $\mu(x,y)$ by
\begin{equation}
\label{eq:mob2}
f(x) = \sum_{y\leq x}\mu(y,x)g(y)
\end{equation}
where $\mu$ is defined inductively by
\[
\mu(x,y) = \left\{      \begin{array}{ll}
                        1, & \text{ if } x=y\\
                        -\sum_{x\leq t< y}\mu(x,t), & \text{ if } x< y\\
                        0, &  \text{ otherwise}.
                        \end{array}     \right.
\]
More precisely, $\mu$ is obtained as the inverse of the Riemann function
$\zeta(x,y):=1$ if $x\leq y$ and 0 otherwise, in the sense that
$\zeta\star\mu=\delta$, where $\star $ is a group operation on real functions on
$X^2$ defined by:
\[
(f\star    g)(x,y) = \sum_{x\leq u\leq y}f(x,u)g(u,y), \quad x,y\in X,
\]  
and $\delta(x,y)=1$ iff $x=y$ and 0 otherwise, is the neutral element.

Viewing in equation (\ref{eq:mob1}) $g$ as the primitive function of $f$, we
can say that in some sense $f$ is the derivative of $g$. Hence, $\mu(x,y)$ acts
as a differential operator. 

In the sequel, our main interest will be capacities and set functions, so that
the partially ordered set is the Boolean lattice of subsets of a finite set
$N$, and $f,g$ are real-valued set functions, or more restrictively
capacities. In this case, for any $A\subseteq B\subseteq N$ we have $\mu(A,B)
= (-1)^{|B\setminus A|}$, and denoting set functions by $v,m$, formulas
(\ref{eq:mob1}) and (\ref{eq:mob2}) become
\begin{align}
v(A) = & \sum_{B\subseteq A} m(B) \label{eq:mob11}\\
m(A) = & \sum_{B\subseteq A}(-1)^{|A\setminus B|}v(B).\label{eq:mob21}
\end{align}
The set function $m$ is called the \emph{M\"obius transform} of $v$. When
necessary, we write $m^v$ to stress the fact it is the M\"obius transform of
$v$. In
cooperative game theory, $m$ is called the \emph{dividend} of the game $v$
\cite{har63,owe88}. In the field of decision theory, $v$ is a capacity and its
M\"obius transform is a fundamental concept (see e.g. Shafer \cite{sha76},
Chateauneuf and Jaffray \cite{chja89}, Grabisch \cite{gra96f}). 

\subsection{The ordinal M\"obius transform}
Let $(L,\leq)$ be a linear reflection
lattice, with $L^+$ its positive part. Consider two $L$-valued functions on
$X$, denoted by $f,g$, and the formula:
\begin{equation}
\label{eq:ordmob1}
g(x) = \langle\svee_{y\leq x} f(y)\rangle.
\end{equation}
To enforce uniqueness of this expression, we use some rule of computation. 
By analogy with the classical case, any solution $f$ to the above equation
plays the role of an  \emph{ordinal} M\"obius transform of $g$, defined with
respect to the given rule of computation. 

Contrary to the classical case, there is not always a solution to this
equation, and if there is one, it may be not unique. Consider the following
example: take $X=\{a,b\}$ with  $a<b$, and $g(a)=\1$,
$g(b)=-\1$. We necessarily have $f(a)=g(a)=\1$, and
$g(b)=f(b)\svee f(a)$. But this last equation reads $-\1=f(b)\svee \1$, which
is impossible to satisfy. Let us put now $g(b)=\1$. Then any $f$ such that
$f(a)=\1$ and $f(b)\neq -\1$ is a solution. 

The following result shows that, at least for the splitting rule, $g$ should
satisfy some properties.
\begin{prop}
\label{prop:sol}
If Equation (\ref{eq:ordmob1}) has a solution for the splitting rule
$\langle\cdot\rangle_-^+$, then necessarily $g$ fulfills
\[
\forall x\in X,\begin{cases}
                |g(x)|\geq|g(y)|, & \forall y\prec x\\
                \text{or} & \\
                g(x)=\0. &
                \end{cases}\tag{*}
\]
\end{prop}
\begin{pf}
Suppose (*) does not hold. Then there exists some $x\in X$ such that
$g(x)\neq\0$ and $|g(x)|<|g(y_0)|$ for some $y_0\prec x$. We have, assuming $f$
is a solution of  (\ref{eq:ordmob1}),
\begin{align*}
g(x) = & \langle \svee_{y\leq x} f(y)\rangle_-^+\\
        = & \langle \svee_{y\leq y_0} f(y)\svee\svee_{\substack{y\leq
        x\\y\not\in[0,y_0]}}f(y)\rangle_-^+ 
\end{align*}
Applying Prop. \ref{prop:cr2} (iii), we get:
\[
|g(x)|\geq|\langle \svee_{y\leq y_0} f(y)\rangle_-^+|=|g(y_0)| \text{ or } g(x)=\0,
\]
which contradicts the hypothesis, hence $f$ is not a solution. 
\qed
\end{pf}

In this section, assuming $|g|$ is isotone (hence fulfilling conditions of Prop.
\ref{prop:sol}), we will give solutions to this equation for a subset of
$\mathfrak{R}$, which are expressed through the inverse of the Riemann function
as in the classical case. Other solutions may exist, but their detailed study is
beyond our scope.

We begin by some considerations close to the classical case. We consider the
following set of functions:
\[
\mathcal{G} = \{f:X^2\longrightarrow L|f(x,x)=\1, \quad f(x,y)=\0 \text{
if } x>y\}, 
\] 
equipped with the following operation $\circledast   $ internal on
$\mathcal{G}$:
\[
(f\circledast    g)(x,y) := \langle\svee_{x\leq u\leq y}[f(x,u)\swedge
g(u,y)]\rangle,
\]
with the same computation rule as in (\ref{eq:ordmob1}). 
The $\circledast$ operation can be defined also when one of the function has
domain $X$: $(f\circledast    g)(x,y) := \langle\svee_{x\leq u\leq y}[f(u)\swedge
g(u,y)]\rangle$. 
Contrary to the classical case, $(\mathcal{G},\circledast )$ has not the
structure of a group. The lack of distributivity in $(L,\svee,\swedge)$ forbids
the satisfaction of associativity in $(\mathcal{G},\circledast )$. However, a
neutral element always exists, and is defined by
\[
\delta(x,y):= \left\{   \begin{array}{ll}
                        \1, & \text{ if } x=y\\
                        \0, & \text{ otherwise}
                        \end{array} \right.
\]
as it is easy to check. Left and right inverses of $f$ may exist and are
not unique in general. Specifically, the left inverse $f^{-1}$ should satisfy:
\[
\langle\svee_{x\leq u\leq y}[f^{-1}(x,u)\swedge f(u,y)]\rangle= \left\{
                                                        \begin{array}{ll}
                                                        \1, & \text{ if } x=y\\
                                                        \0, & \text{ otherwise}
                                                        \end{array} \right.
\]
from which we deduce that
\begin{align}
f^{-1}(x,x)= & \1, \quad \forall x\in X\label{eq:inv1}\\
\langle\svee_{x\leq u\leq y}[f^{-1}(x,u)\swedge f(u,y)]\rangle= & \0, \quad
\forall x<y.\label{eq:inv2} 
\end{align}
Defining $f^{-1}(x,y)=\0$ whenever $x>y$ and using (\ref{eq:inv1}), we
deduce that $f^{-1}$ belongs to $\mathcal{G}$.  The following lemma clarifies
the situation for the Riemann function $\zeta(x,y)$.
\begin{lem}
\label{lem:3}
The inverse of the Riemann function (left or right) is given by
\begin{align*}
\zeta^{-1}(x,x) = & \1, \quad\forall x\in X\\
\zeta^{-1}(x,y) = & -\1, \quad\forall x,y\in X \text{ such that } x\prec y\\
\end{align*}
for all $\langle\cdot\rangle\in\mathfrak{R}$, and for $x,y$ such that $x<y$ and
$x\not\prec y$  
\begin{itemize}
\item For any rule in $[\langle\cdot\rangle_-^+,\langle\cdot\rangle_=]$,
$-\1, \0$ and $\1$ are possible values for $ \zeta^{-1}(x,y)$. In particular,
if $\langle\cdot\rangle=\langle\cdot\rangle_=$, these are the only possible
values, and if $\langle\cdot\rangle=\langle\cdot\rangle_-^+$, all values in $L$
are possible.
\item There exists no inverse in general for any rule in
$]\langle\cdot\rangle_=,\langle\cdot\rangle_0]$. If $X$ is linearly ordered,
then $\zeta^{-1}(x,y)=\0$ is solution for any rule in $\mathfrak{R}$.
\end{itemize}
\end{lem}
\begin{pf}
We know already from (\ref{eq:inv1}) that $\zeta^{-1}(x,x) = \1$ for any
computation rule.  Equation (\ref{eq:inv2}) for the Riemann function becomes
\[
\langle \svee_{x\leq u \leq y}\zeta^{-1}(x,u)\rangle = \0, \quad \forall x<y.
\]
If $x\prec y$, then clearly we get $\zeta^{-1}(x,y) = -\1$ as only solution,
and for any computation rule. Let us consider $x,y$ such that $x\prec u\prec
y$. The above equation reads
\[
\langle \1\svee\svee_{u\prec y}(-\1)\svee\zeta^{-1}(x,y)\rangle =\0.
\]
Note that it suffices to show that the above sequence of terms belongs to
$\mathcal{O}_{\langle\cdot\rangle}$. 
In the case of the splitting rule $\langle\cdot\rangle_-^+$, clearly any number
in $L$ is solution for $\zeta^{-1}(x,y)$. In the case of the weak rule
$\langle\cdot\rangle_=$, only $\1,\0,-\1$ are solutions. Then for any rule in $
[\langle\cdot\rangle_-^+,\langle\cdot\rangle_=]$, the result is proven using
Prop. \ref{prop:cr} (ii). 
 
Let us consider any rule $\langle\cdot\rangle$ in
$]\langle\cdot\rangle_=,\langle\cdot\rangle_0]$. Then there exist some
sequences in $\mathfrak{S}$ for which the index set of deleted terms is strictly
included in $I_=$. This means that it may exist a poset $(X,\leq)$ such that
the above equation has no solution. Indeed, if $\zeta^{-1}(x,y)=\1$ or $-\1$,
the sequence of $\1,-\1$ we obtain may be such that the index set of deleted
terms is strictly included in $I_=$, and so the result cannot be $\0$. The same
holds if $\zeta^{-1}(x,y)$ takes any other value. In particular, in the case of
the strong rule $\langle\cdot\rangle_0$, observe that if there is a unique
element $u$ between $x$ and $y$, then $\zeta_{-1}(x,y)=\0$ is solution (and due
to Prop. \ref{prop:cr} (ii) (iii), the results extends to any other rule). If
there are two elements $u$ between $x$ and $y$, then $\zeta_{-1}(x,y)=\1$ is
solution. Otherwise, there is no solution.
\qed
\end{pf}

We call \emph{canonical inverse} the solution where $\zeta^{-1}(x,y)=\0$ when
$x<y$ but $x\not\prec y$. It is a solution for all rules in
$[\langle\cdot\rangle_-^+,\langle\cdot\rangle_=]$ (and for any rule in
$\mathfrak{R}$, if $X$ is linearly ordered). By extension, we call it
\emph{canonical pseudo-inverse} for rules outside
$[\langle\cdot\rangle_-^+,\langle\cdot\rangle_=]$, when $X$ is not a linear
order. In the sequel we examine under what conditions inverses of the Riemann
function permit to build solutions.

If $(\mathcal{G},\circledast )$ were a group, then $g\circledast \zeta^{-1}$
should be solution to the equation. Let us study when $f=g\circledast
\zeta^{-1}$ is indeed a solution. The following is the main result of the
paper.
\begin{thm}
\label{th:ordmob}
Assume $g$ is such that $|g|$ is isotone. Then $g\circledast \zeta^{-1}$ is
solution to Equation (\ref{eq:ordmob1}) for any rule in
$[\langle\cdot\rangle^+_-,\langle\cdot\rangle_=]$, where $\zeta^{-1}$ is any
inverse of the Riemann function. For rules in
$]\langle\cdot\rangle_=,\langle\cdot\rangle_0]$ and the canonical
pseudo-inverse, $g\circledast \zeta^{-1}$ is not a solution in general.
\end{thm}
(see proof in Appendix)

Equation (\ref{eq:ordmob1}) may have no solution at all for the strong rule,
even if $|g|$ is isotone. Indeed, take $X=\{0,a,b,c\}$ with $0\prec a\prec c$
and $0\prec b\prec$, $b,c$ being incomparable, and define $g(0)=\0$,
$g(a)=g(b)=-\1$, and $g(c)=\1$. Then clearly $f(0)=\0$, $f(b)=f(c)=-\1$ and
there is no solution for $f(c)$.

The preceding results can be summarized as follows.
\begin{summ}
We consider $f,g:X\longrightarrow L$, and the following equation to solve:
\[
g(x) = \langle\svee_{y\leq x}f(y)\rangle
\]
with
$\langle\cdot\rangle\in[\langle\cdot\rangle^+_-,\langle\cdot\rangle_=]$. We
call \emph{M\"obius function} $\mu(x,y)$ any inverse $\zeta^{-1}$ of the
Riemann function, as given in Lemma \ref{lem:3}, and call \emph{canonical
M\"obius function} the canonical inverse of the Riemann function.

Assuming that $|g|$ is isotone, then $f(x)=(g\circledast \mu)(x)$ is solution
for any M\"obius function, where $\circledast$ is defined with respect to the
corresponding computation rule. We call any such $f$ a \emph{M\"obius
transform} of $g$, and \emph{canonical M\"obius transform of $g$}, denoted
$m^g$, the one corresponding to the canonical M\"obius function. It is given by:
\begin{equation}
\label{eq:canonmob}
 m^g(x) := \langle g(x)\svee\Big[-\svee_{y\prec x}g(y)\Big]\rangle.
\end{equation}
\end{summ}

\subsection{The case of non negative isotone functions}
A particular case of interest is to restrict to isotone functions valued on
$L^+$ (capacities correspond to this case, hence its interest). Let us call
them non negative isotone functions.
\begin{thm}
\label{th:ordmob2}
For any non negative isotone function $g$, the set of non negative solutions to
 Equation (\ref{eq:ordmob1}) is the interval $[m_*,m^*]$, defined by:
\begin{align*}
m^*(x) = & g(x), \quad\forall x\in X\\
m_*(x) = m^g(x) = & \left\{     \begin{array}{ll}
                        g(x), & \text{ if } g(x)>g(y), \quad \forall y\prec x\\
                        \0, & \text{ otherwise}
                        \end{array}
                \right., \quad\forall x\in X.
\end{align*} 
\end{thm}
\begin{pf}
Since $g$ is isotone and non negative, $m^*$ is clearly a solution. On the
other hand, Th. \ref{th:ordmob} applies, and we recognize $m_*$ as the
canonical M\"obius transform (\ref{eq:canonmob}).

We have to prove that these are indeed the lower and upper bounds of non
negative solutions. If $m^*$ were not the upper bound, it should exist $x_0\in
X$ such that $m^*(x_0)>g(x_0)$. Then due to isotonicity, we would have
$g(x_0)<\bigvee_{y\leq x_0} m^*(y)$, a contradiction. Similarly, if
$m_*(x_0)<g(x_0)$ for some $x_0$ such that $g(x_0)>g(y)>\0$ for all $y\prec
x_0$, we would have $g(x_0)>\bigvee_{y\leq x}m_*(y)$, a contradiction again.

Lastly, we show that any
$f\in[m_*,m^*]$ is also a solution. Since $m_*,m^*$ are non negative solutions,
we have for any $x$ 
\[
\bigvee_{y\leq x} m_*(y) = \bigvee_{y\leq x} m^*(y).
\]
Since $\bigvee$ is increasing, any $m\in[m_*,m^*]$ will also satisfy the
equation. 
\qed
\end{pf}

In case of no ambiguity, we denote simply $m^g$ by $m$.
 Moreover, since our framework is ordinal in the rest of the paper,
we will omit to call it ``ordinal'', and will use the term ``classical''
M\"obius transform when referring to the usual definition. We denote by $[m]$
the interval $[m_*, m^*]$, and with some abuse of notation, any function in
this interval.

Some remarks are of interest at this stage.
\begin{rem}
As with the classical case, the M\"obius transform has the meaning of a
derivative. From Definition \ref{def:deri} and (\ref{eq:deriv}), it is clear
that $m\equiv g'$.
\end{rem}
\begin{rem}
Since $f,g$ are non negative, we need no more computation rules in (\ref{eq:ordmob1}). However,
negative solutions exist. It is easy to check that for any computation rule,
$m_*$ can be defined by
\[
m_*(x) =  \left\{       \begin{array}{ll}
                        g(x), & \text{ if } g(x)>g(y), \quad \forall y\prec
                        x\\
                        \text{any }e\in L,e\succ -g(x),  & \text{ otherwise}
                        \end{array}
                \right.,
\]
$\forall x\in X$. However, negative solutions do not possess good properties,
and would not permit to obtain the subsequent results.
\end{rem}

\begin{defn}
Let $g$ be any isotone function from $X$ to $L^+$. We call \emph{$g$-chain} any
chain $C$ in $X$ such that $g(x)$ is constant on $C$, and there is no chain
$C'\supset C$ such that $g(x)$ is constant on $C'$. The set of all
$g$-chains is denoted $\mathscr{C}(g)$. The \emph{value} of a $g$-chain $C$ is
defined by $g(C):=g(x)$ for some $x\in C$.
\end{defn}
Any $g$-chain $C$ has a unique minimal element, denoted $C_*$. Indeed, either
$C$ is finite or infinite. In the first case, the results trivially hold. In
the second case, since 0 is the unique minimal element of $X$, and $X$ is
locally finite, $C$ has the form $\{x|x\geq a\}$, hence the result. On the
contrary, there is not always a maximal element $C^*$.

If a $g$-chain $C$ is finite, its length
is defined as usual by $l(C):= |C|-1$.

The following is easy to show (proof is omitted).
\begin{prop}
\label{prop:chain}
Let $g$  be any isotone function from $X$ to $L^+$, and $C$ be any
$g$-chain. Then:
\begin{itemize}
\item[(i)] $\mathscr{C}(g)=\emptyset$ iff $m\equiv g$.
\item [(ii)] If $(X,\leq)=(2^N,\subseteq)$ where $N$ is a finite set of $n$
elements,  and $g(\emptyset)< g(N)$, then $l(C)<n$ (i.e. $C$ is not a maximal
chain), 
for any $C\in \mathscr{C}(g)$.
\item [(iii)] Let $\mathscr{C}(g)\neq\emptyset$ and $C\in\mathscr{C}(g)$. Then
        \begin{itemize}
        \item [(iii.1)] For all $x\in C$, $x\neq C_*$, $m(x)=\0$.
        \item [(iii.2)] $m(C_*)=g(C_*)$.
        \item [(iii.3)] For all $x\not\in C, \forall C\in \mathscr{C}(g)$, $m(x)
= g(x)$. 
        \end{itemize}
\end{itemize}
\end{prop}

Let us suppose now that $X$ and $L^+$ are endowed with a conjugation mapping
$\overline{\cdot}$. Then necessarily, $X$ has a unique maximal element, denoted 1,
and any $g$-chain is finite, with a unique maximal and minimal element. We
define the \emph{conjugate of $g$} by $\overline{g}(x)
:=\overline{g(\overline{x})}$. In this section, we compute $m^{\overline{g}}$
and express it with respect to $m^g$. The following can be shown.
\begin{prop}
\label{prop:conju}
Under the above assumptions, let $\overline{g}$ be the conjugate function of
$g:X\longrightarrow L^+$. Then:
\begin{itemize}
\item[(i)] the set of $\overline{g}$-chains is given by
\[
\mathscr{C}(\overline{g}) =
\{\overline{C}:=\{\overline{c_l},\ldots,\overline{c_1}\}|\{c_1,\ldots,c_l\}=:C,
C\in \mathscr{C}(g)\}.
\]
and $\overline{g}(\overline{C}) = \overline{g(c)}$.
\item [(ii)] the M\"obius transform of $\overline{g}$ is given by 
\begin{equation}
m^{\overline{g}}(x) = \left\{ \begin{array}{ll}
                        \0, & \text{ for all }x\text{ in some }\overline{C}\in
                        \mathscr{C}(\overline{g}), x\neq \overline{C}_*\\
                        \overline{m^g(C_*)}, & \text{ if } x=\overline{C}_*\\
                        \overline{m^g(\overline{x})}, & \text{ otherwise}.
                        \end{array}
                \right.
\end{equation}
\end{itemize}
\end{prop}
\begin{pf}
(i) Let us consider $C\in
\mathscr{C}(g)$, and $C=\{c_1,\ldots,c_l\}$ with $c_1<\cdots<c_l$. Since $g$ is
constant over $C$, we get
$\overline{g}(\overline{c_1})=\cdots=\overline{g}(\overline{c_l})$,
$\overline{c_l}<\cdots<\overline{c_1}$, which means that
$\overline{C}:=\{\overline{c_l},\ldots,\overline{c_1}\}$ is a
$\overline{g}$-chain. Also, we have $g(C) = \overline{g(\overline{C})}$.

(ii)  Suppose that
$\mathscr{C}(g)=\emptyset$. Then $\mathscr{C}(\overline{g})=\emptyset$ too, and
due to Prop. \ref{prop:chain} (i), $m^{\overline{g}}\equiv\overline{g}$. This
leads to
\[
m^{\overline{g}}(x) = \overline{m^g(\overline{x})}.
\]
Suppose now that $\mathscr{C}(g)\neq\emptyset$, and $C\in \mathscr{C}(g)$, with
corresponding $\overline{C}\in \mathscr{C}(\overline{g})$. By
Prop. \ref{prop:chain} (iii), if $c\in
\overline{C}$, $c\neq\overline{C}_*$, then $m^{\overline{g}}(c)=\0$. If
$c=\overline{C}_*$, then $m^{\overline{g}}(c)= \overline{g}(c) =
\overline{g(\overline{c})}$. Remark that $\overline{c}=C^*$, so that
$m^g(\overline{c})=\0$. But $m^g(C_*) = g(C_*) = g(\overline{c})$ (since
$\overline{c}$ and $C_*$ are in $C$), hence the result.
\qed
\end{pf}

It is possible to have a slightly more compact form for this result. 
 Let us denote by $n_C(c)$ the element in
$C$ which has the symmetric place of $c$ (i.e. $n_C(c_k)= c_{l-k+1}$). We can
write for $c=\overline{C}_*$:
\[
m^{\overline{g}}(c) = \overline{m^g(n_C(\overline{c}))}
\]
(see figure \ref{fig:conju} below).
Considering that any $c$ not belonging to a $g$-chain is itself a chain $C$ of
0 length, so that $n_C(c)=c$, we have the general result:
\begin{equation}
m^{\overline{g}}(x) = \left\{ \begin{array}{ll}
                        \0, & \text{ for all }x\text{ in some }\overline{C}\in
                        \mathscr{C}(\overline{g}), x\neq \overline{C}_*\\
                        \overline{m^g(n_C(\overline{x}))}, & \text{otherwise}.
                        \end{array}
                \right.
\end{equation}
\begin{figure}[htb]
\begin{center}
$
\epsfysize=5cm
\epsfbox{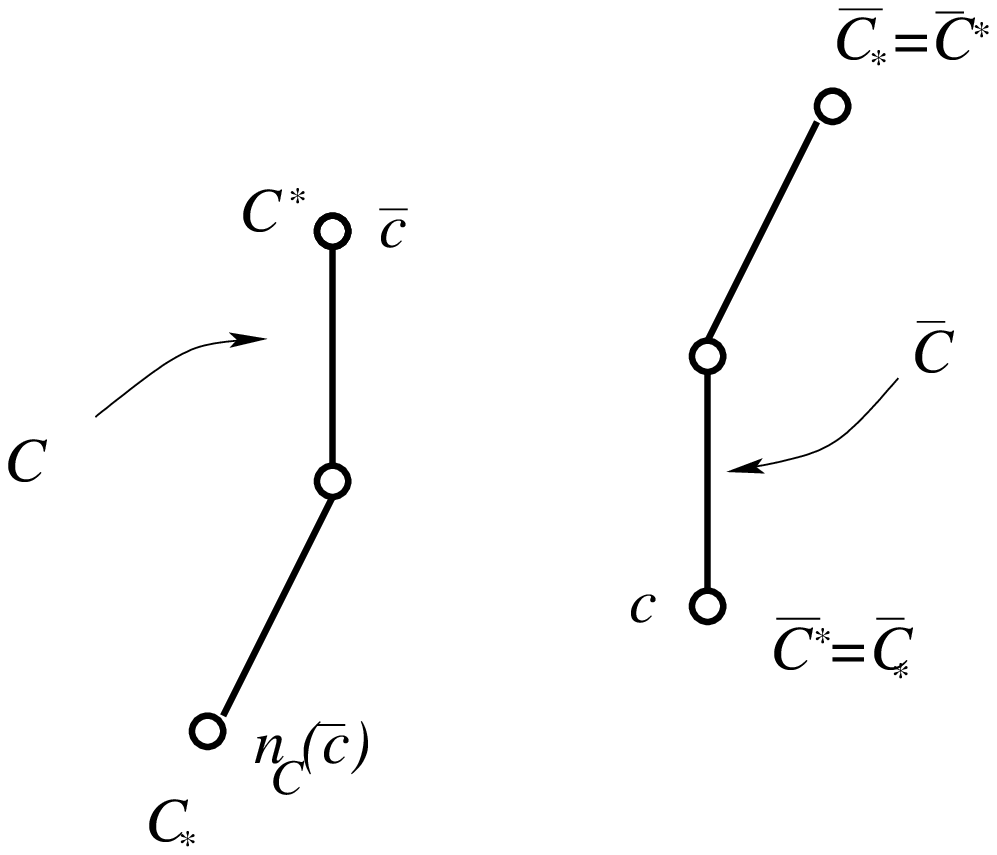}
$
\end{center}
\caption{$g$-chains and $\overline{g}$-chains}
\label{fig:conju}
\end{figure}

The classical M\"obius transform can be viewed as a linear operator on the set
of real functions on $X$. We may expect that the ordinal counterpart has
a similar property with $\svee,\swedge$, i.e. $m^{f\svee g}=m^f\svee m^g$ and
$m^{\alpha\swedge f} = \alpha\swedge m^f$, for any $\alpha\in L^+$. However,
the following simple example shows that this is not the case. 
\begin{exmp}
Let us take $X$ to be the Boolean lattice $2^2$ whose elements are denoted
$\emptyset,\{1\},\{2\},\{1,2\}$, and consider two functions $g_1,g_2$ defined
as follows:
\begin{center}
\begin{tabular}{|c|c|c|c|c|}\hline
        & $\emptyset$ & $\{1\}$ &  $\{2\}$ & $\{1,2\}$ \\ \hline
$g_1$   & $\0$ & $\0$ &$\0$ &$\1$ \\ \hline
$g_2$   & $\0$ & $\1$ &$\1$ &$\1$ \\ \hline 
\end{tabular}
\end{center}
The computation of the M\"obius transform $m_*$ gives
\begin{center}
\begin{tabular}{|c|c|c|c|c|}\hline
        & $\emptyset$ & $\{1\}$ &  $\{2\}$ & $\{1,2\}$ \\ \hline
$m_*[g_1]$      & $\0$ & $\0$ &$\0$ &$\1$ \\ \hline
$m_*[g_2]$      & $\0$ & $\1$ &$\1$ &$\0$ \\ \hline 
\end{tabular}
\end{center}
Clearly, $g_1\svee g_2 = g_2$, but $m_*^{g_1}\svee m_*^{g_2} \neq
m_*^{g_2}$. 
\end{exmp}
Remarking that $m^*$ is maxitive, one should expect that it is possible to find
some $m\in [m]$, $m<m^*$ at least on some element of $X$. The above example
shows that this is even impossible in general: due to the fact that
$m^{g_1}(\{1,2\})=\1$, we must have $m^{g_2}(\{1,2\})=\1$, and thus $m\equiv
m^*$.

\section{The ordinal M\"obius transform of capacities}
We devote this section to the particular case of capacities on some finite set
$N:=\{1,\ldots,n\}$, which is our original motivation in this paper.  Then
$X=2^N$ is a Boolean lattice, and we suppose in addition that $L^+$ is a
conjugation linear lattice. Capacities are denoted by $v$.

A first fact is that we can give an alternative expression of the (canonical)
M\"obius transform, which is very similar to the classical one
(\ref{eq:mob21}).
\begin{equation}
m(A) := \bigvee_{B\subseteq A, |A\setminus B|
\mbox{\scriptsize  even}}v(B)
\svee\left(- \bigvee_{B\subseteq A, |A\setminus B| \mbox{\scriptsize odd}}v(B)\right)
\end{equation}
for any $A\subseteq N$. Indeed,
\[
\bigvee_{B\subseteq A, |A\setminus B|
\mbox{\scriptsize  even}}v(B) = v(A)
\]
and 
\[
\bigvee_{B\subseteq A, |A\setminus B| \mbox{\scriptsize odd}}v(B) =
\bigvee_{B\prec A}v(B)
\]
so that we recognize (\ref{eq:canonmob}).

\begin{figure}[htb]
\begin{center}
$
\epsfxsize=14cm
\epsfbox{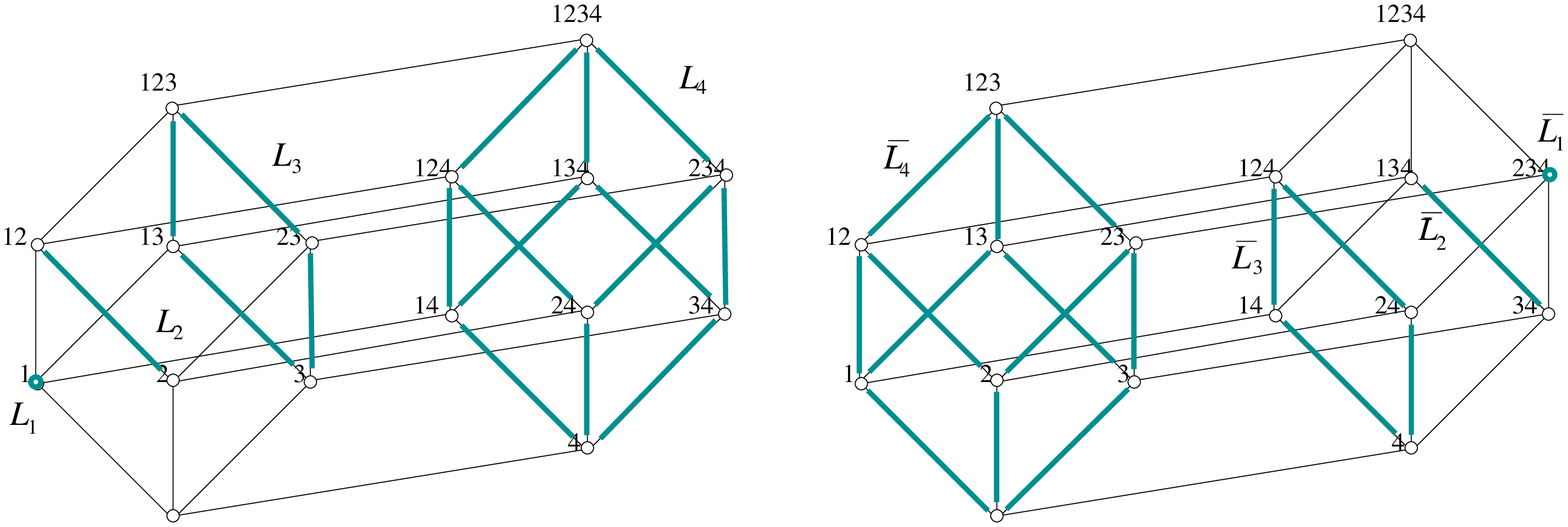}
$
\end{center}
\caption{$\Pi$-chains (left) and N-chains (right) with $N=\{1,2,3,4\}$.}
\label{fig:latti}
\end{figure}

In the field of decision theory and artificial intelligence,
sup-preserving functions from $X$ to $L^+$ (i.e. such that $g(x\vee y) =
g(x)\vee g(y)$, for every $x,y\in X$) are called \emph{possibility measures}
\cite{zad78,dupr80} or \emph{maxitive measures}, and are denoted by $\Pi$. By
conjugation we have $\overline{g}(x\wedge y) = \overline{g}(x)\wedge
\overline{g}(y)$, for every $x,y\in X$ (inf-preserving functions), they are
called \emph{necessity measures} or \emph{minitive measures}, and are denoted
by $\mathrm{N}$. Remark that for any
$A=\{i_1,\ldots,i_l\}\subseteq N$, we have $\Pi(A) = \bigvee_{i\in
A}\Pi(\{i\})$. The following can be shown.
\begin{thm}
Let $\Pi,\mathrm{N}$ be a pair of conjugate possibility and necessity measures, and
suppose without loss of generality that the elements in $N$ are such that
$\Pi(\{1\})\leq\cdots\leq\Pi(\{n\})$. Then
\begin{itemize}
\item the M\"obius transform of $\Pi$ is non zero on an antichain:
\[
m^\Pi(A) = \left\{      \begin{array}{ll}
                        \Pi(\{i\}), & \text{ if } A=\{i\}, i\in N\\
                        \0, & \text{ otherwise.}
                        \end{array} \right.
\]
\item the M\"obius transform of $\mathrm{N}$ is non zero on a chain. Assuming
$\0< \Pi(\{1\})<\cdots<\Pi(\{n\})=\1$, the expression is:
\[
m^{\mathrm{N}}(A) = \left\{     \begin{array}{ll}
                        \overline{\Pi(\{i\})}, & \text{ if }
                        A=\{i+1,\ldots,n\}, i\in N\\
                        \0, & \text{ otherwise.} 
                        \end{array} \right.
\]
If $\Pi(\{i\}) = \Pi(\{i+1\})$ for some $i$, then
$m^{\mathrm{N}}(\{i+1,\ldots,n\})=\0$. 
\end{itemize}
\end{thm}
\begin{pf}  Let us  suppose $\0<
\Pi(\{1\})<\cdots<\Pi(\{n\})=\1$.

(i) Let us compute
$\mathscr{C}(\Pi)$. Let us consider $i\in N$. We denote by
$L_i$ the sublattice which is the interval $[\{i\},\{1,\ldots,i\}]$. By
construction, any subset $A\in L_i$ is such that $\Pi(A)=\Pi(\{i\})$, and only
those ones, which proves that all $\Pi$-chains with value $\Pi(\{i\})$ are the
maximal chains of $L_i$. In other words, $\mathscr{G}(\Pi(\{i\}))=L_i$. Now,
the bottom of $L_i$ being $\{i\}$, we get the result. 

(ii) From Prop. \ref{prop:conju} (i), we know that $\mathscr{C}(\mathrm{N})$ is
in some sense the symmetric of $\mathscr{C}(\Pi)$ in the lattice $2^N$. More
precisely, the sublattices of interest are
$\overline{L}_i:=[\overline{\{1,\ldots,i\}},\overline{\{i\}}]$. They correspond
to the groups $\mathscr{G}(\overline{\Pi(\{i\})})$, and since the bottom
element of $\overline{L}_i$ is $\overline{\{1,\ldots,i\}}$, and  only $N$ does
not belong to any $\overline{L}_i$, we get the desired
result.

If $\Pi(\{i\}) = \Pi(\{i+1\})$ for some $i$, then it is easy to check that
subset $\{i+1,\ldots,n\}$ disappears in the chain, but there is no change for
$\Pi$. 
\qed
\end{pf}
Figure \ref{fig:latti} illustrates the result. 

\section{Applications of symmetric ordered structures and perspectives}
We conclude the paper by indicating several possible applications of our
symmetric ordered structure. We mainly developed in this paper the theory of
M\"obius transform, and its application to capacities. We briefly mentioned in
the introduction that one of the main motivation was the definition of a
symmetric Sugeno integral. Clearly, our aim is achieved, since Equation
(\ref{eq:symsug}) is now perfectly defined, and could be a starting point to
develop an ordinal or \emph{qualitative} counterpart of Cumulative Prospect
Theory, a theory which is of primary importance in e.g. economics. We refer the
reader to \cite{gra01e} for a detailed study of the symmetric Sugeno integral,
along with other results on capacities and the ordinal M\"obius
transform. Based on symmetric ordered structures and the symmetric Sugeno
integral, we have already built a general model of multicriteria decision
making \cite{grdila01}, which permits to tackle real problems where only
qualitative information is available. This is indeed a common situation in
many applications (e.g. project selection, subjective evaluation of consumer
goods, etc.). 

Another application would be to investigate capacities
defined on arbitrary lattices instead of the usual Boolean lattice
\cite{grla03c}, a new promising topic in decision making. Considering these
general capacities, valued on $L$ instead of a
real interval, we need  our general results from Sections \ref{sec:nonass} and
\ref{sec:motr} to get the M\"obius transform and properly  define a general
Sugeno integral.  

\medskip

On a purely mathematical point of view, we have studied in detail algebraic
properties of our new structure, and in particular the possible ways to escape
from non associativity. The generality of our results may open new areas related
to ordered structures and combinatorics. It might also be viewed as a starting
point of ordinal ``linear'' algebra, noticing that $\circledast$ is in fact the
matrix product. We describe hereafter a possible application of this ordinal
linear algebra. Considering two finite universal sets $X,Y$, a \emph{fuzzy
  binary relation} or {valued binary relation }on $X\times Y$ is simply a
function $R:X\times Y\longrightarrow [0,1]$, where $R(x,y)$ is the strength of
relation between $x$ and $y$. Many results exist in this area (see e.g.
\cite{nosepesa89,foro94}), but we are interested here in what is called
\emph{fuzzy relation equations}, which are important in system theory.
Considering finite universal sets $X,Y,Z$ and three fuzzy relations $P,Q,R$ on
$X\times Y, Y\times Z, X\times Z$ respectively, we consider the equation
$R=P\circ Q$, which we want to solve for $P$. Composition of relations is given
by:
\[
R(x,z)=\bigvee_{y\in Y}(P(x,y)\wedge R(y,z)).
\]  
The solution set of this equation, whenever non empty, has the structure of a
union of intervals $[\check{P}_i,\hat{P}]$, where $\hat{P}$ is the unique
maximal solution, and $\check{P}_i$ are minimal ones. Allowing fuzzy relations
to be valued in $[-1,1]$ or any symmetric linear order (\emph{bipolar} fuzzy
relation), replacing $\vee,\wedge$ by $\svee,\swedge$ and considering a
particular computation rule, the above equation coincides with our
$\circledast$ operation. Hence our results could provide powerful tools for
solving bipolar fuzzy relation equations, a topic which has never been
addressed, but which may become important in the near future, since bipolar
scales deserve a great interest in this field. 

An interesting further study would be to change the starting point,
e.g. enforcing associativity and loosing symmetry. We already know from
Prop.~\ref{prop:best} (2) some properties of this kind of structure.

\begin{ack}
This work has benefited from many fruitful discussions with 
D. Denneberg, J.L. Marichal, and T. Murofushi, who are deeply acknowledged.
The author thanks also the anonymous referees for their constructive comments. 
\end{ack}

\bibliographystyle{plain}
\bibliography{../BIB/fuzzy,../BIB/grabisch,../BIB/general}

\appendix

\section{Proof of Theorem \ref{th:ordmob}}
We first need a technical lemma.
\begin{lem}
\label{lem:4}
Let $f=g\circledast \zeta^{-1}$, where $\zeta^{-1}$ is any
inverse of the Riemann function, and assume that $|g|$ is isotone. Then, for
any $x\in X$:
\begin{itemize}
\item [(i)] If for all $ y\prec x$, $|g(x)|>|g(y)|$  or $g(x)=-g(y)$,
\[
f(x) = g(x)
\]
for any computation rule in $\mathfrak{R}$.
\item [(ii)] If there exists $y\prec x$ such that $ g(x)=g(y)$
        \begin{itemize}
        \item [(ii.1)] in the case of the splitting rule
$\langle\cdot\rangle_-^+$, then $ f(x)= \0$. 
        \item [(ii.2)] for any rule in
$[\langle\cdot\rangle^+_-,\langle\cdot\rangle_=]$,
we have $|f(x)|  < |g(x)|$.
        \end{itemize}
\item [(iii)] For any rule in $[\langle\cdot\rangle_=,\langle\cdot\rangle_0]$,
and for the canonical pseudo-inverse 
\[
f(x)=\begin{cases}
        \0, & \text{ if  } |G^+|=|G^-|+1  \\
        \text{either } g(x), \0 \text{ or } -g(x) & \text{ otherwise }
        \end{cases}
\]
with $G^+:=\{y\in X\,;\, y\prec x \text{ and } g(y)=g(x)\}$, and $G^-:=\{y\in
X\,;\, y\prec x \text{ and } g(y)=-g(x)\}$ . In the case of the strong rule
$\langle\cdot\rangle_0$, the result particularizes as follows
\[
f(x)=\begin{cases}
        \0, & \text{ if  } |G^+|=|G^-|+1  \\
        g(x), & \text{ if }|G^+|\leq|G^-| \\
        -g(x), & \text{ otherwise.}
        \end{cases}
\]
\item [(iv)] if $y\leq x$ and $|g(y)|<|g(x)|$, then $|f(y)|<|g(x)|$ for all
computation rules in $\mathfrak{R}$.
\end{itemize}
\end{lem} 
\begin{pf}
We have
\begin{align}
f(x) = & g\circledast\zeta^{-1}(x)= \langle \svee_{u\leq
x}g(u)\swedge\zeta^{-1}(u,x)\rangle\nonumber \\
        & = \langle g(x)\svee\svee_{u\prec x}(-g(u))\svee\svee_{u\prec v\prec x}(g(u)\swedge\zeta^{-1}(u,x))\svee\dots\rangle. \label{eq:*}
\end{align}
Let us remark that 
\begin{equation}
\label{eq:pro}
|g(u)\swedge\zeta^{-1}(u,x)|\leq |g(u)|
\end{equation}
for all $u\leq x$.  If $|g(x)|>|g(y)|$ for all $y<x$ or $g(x)=-g(y)$ for some
$y\prec x$, then by (\ref{eq:pro}) clearly associativity holds in (\ref{eq:*}),
so that for any computation rule the result is the same, which is $g(x)$. This
proves (i).

Suppose there is some $y\prec x$ such that $g(x)=g(y)$. Using (\ref{eq:pro})
and due to the isotonicity of $|g|$, extremal terms in $f(x)$ are $g(x)$ and
$-g(x)$. This proves that $f(x)=\0$ for the splitting rule. If the weak rule is
used, then all terms $g(x),-g(x)$ disappear, so that we can only deduce that
$|f(x)|<|g(x)|$. Now, using Prop. \ref{prop:cr} (i), we have proven (ii).

(iii) is clear since $f(x)=\langle g(x)\svee(-\svee_{y\prec x} g(y))\rangle$.
  (iv) comes from (\ref{eq:pro}), isotonicity of $|g|$, and
  Prop. \ref{prop:cr2} (i).
\qed
\end{pf}

\begin{pf} (Th. \ref{th:ordmob})
We assume that $g(x)\neq\0$, otherwise the result holds trivially.
Let $x\in X$. Assume $|g(x)|>|g(y)|$ or $g(x)=-g(y)$ for all $y\prec x$. Then
by Lemma \ref{lem:4} (i), $f(x)=g(x)$, and by Lemma \ref{lem:4} (iv),
$|f(y)|<|g(x)|$. Hence $\langle\svee_{y\leq x}f(y)\rangle = \langle
f(x)\svee\svee_{y<x}f(y)\rangle = g(x)$ as expected, since associativity holds.

Assume $g(x)=g(y)$ for some $y\prec x$. Let us introduce $C_x:=\{y\in X|g(y)=g(x), y< x\}$. Since
0 is the unique minimal element of $X$, $C_x\subseteq[0,x]$ and hence is
finite. Thus, $C_x$ possesses at least one minimal element. Let us denote by
$C_{x*}$ the set of these minimal elements. We have:
\[
\langle \svee_{y\in C_x} f(y)\rangle =
\langle\svee_{y\in C_{x*}}f(y)\svee\svee_{y\in C_x\setminus C_{x*}}f(y)\rangle.
\]
From Lemma \ref{lem:4} (i), we have $f(y)=g(y)=g(x)$ for all $y\in C_{x*}$, and
for all $y\in C_x\setminus C_{x*}$, we have $|f(y)|<|g(y)|$ for any rule in
$[\langle\cdot\rangle^+_-,\langle\cdot\rangle_=]$ (use Lemma \ref{lem:4} (ii)
and the fact that $|g(x)|>\0$). Hence $\langle \svee_{y\in C_x} f(y)\rangle =
g(x)$ since associativity holds. Now,
\[
\langle\svee_{y\leq x}f(y) \rangle = \langle \svee_{y\in
C_x}f(y)\svee\svee_{\substack{y<x\\y\not\in C_x}}f(y)\rangle.
\]
Since $\svee_{y\in C_x}f(y) = g(x)$ and by Lemma \ref{lem:4} (iv)
$|f(y)|<|g(x)|$ for all $y<x,y\not\in C_x$, we have finally
$\langle\svee_{y\leq x}f(y) \rangle = g(x)$ as desired. 

Consider now a rule in $]\langle\cdot\rangle_=,\langle\cdot\rangle_0]$ and the
canonical pseudo-inverse. From Lemma \ref{lem:4} (iii), we may have
$f(y)=-g(x)$ for some $y\in C_x\setminus C_{x^*}$, so that  $\langle
\svee_{y\in C_x} f(y)\rangle\neq g(x)$ may occur, and the result does not
hold. 
\qed
\end{pf}

\end{document}